\documentclass[a4paper,11pt]{article}
\pdfoutput=1 % if your are submitting a pdflatex (i.e. if you have
\usepackage{jinstpub} % for details on the use of the package, please see the JINST-author-manual
\usepackage{graphicx}
\usepackage{subfig}
\usepackage{lineno}
\usepackage{multirow}
\usepackage{circuitikz}
\title{\boldmath  Operation of a tunable Power over Fiber system for light detectors down to 4.6 K}
\author[a,b]{A.~Andreani}
\author[c,d]{C.~Brizzolari}
\author[c,d]{E.J.~Cristaldo Morales}
\author[c,d]{M.J.~Delgado Gonzalez}
\author[c,d]{A.~Falcone}
\author[a,b,f]{N.~Gallice}
\author[c,d]{C.~Gotti}
\author[a,b]{M.~Lazzaroni}
\author[c,d]{L.~Meazza}
\author[c,d]{G.~Pessina}
\author[e]{D.~Santoro}
\author[c,d]{F.~Terranova}
\author[c,d,1]{M.~Torti, \note{corresponding author}}
\author[a,b]{V.~Trabattoni}

\affiliation[a]{Dipartimento di Fisica, Università degli Studi di Milano, Via G. Celoria 16, 20133 Milano, Italy}

\affiliation[b]{Istituto Nazionale di Fisica Nucleare, Sezione di Milano, Via G. Celoria 16, 20133 Milano, Italy}

\affiliation[c]{Dipartimento di Fisica "G.Occhialini", Università degli Studi di Milano Bicocca, Piazza della Scienza 3, 20126 Milano, Italy}

\affiliation[d]{Istituto Nazionale di Fisica Nucleare, Sezione di Milano Bicocca, Piazza della Scienza 3, 20126 Milano, Italy}

\affiliation[e]{Dipartimento di Ingegneria e Architettura, Università di Parma, Parco Area delle Scienze 181/A, 43124 Parma, Italy}

\affiliation[f]{Brookhaven National Laboratory, PO 5000 Upton, NY 11973, USA}

\emailAdd{marta.torti@mib.infn.it}

\abstract{The Power over Fiber (PoF) technology delivers electrical power by transmitting laser light through a lightweight, non-conductive fiber optic cable to a remote photovoltaic optical converter, which in turn powers sensors or electrical devices.
Among the several advantages offered by this solution are spark-free operation in the presence of electric fields, elimination of noise induced by power lines, immunity to electromagnetic interference, and high robustness in hostile environments.
The R\&D for the application of PoF in cryogenic environments started at FNAL and BNL (USA) in 2020 to power the Photon Detection System of the DUNE Vertical Drift module.
This paper presents the results obtained in the framework of Cryo-PoF project
where we developed a single-laser input line system to power an electronic amplifier and the photosensors at cryogenic temperatures.
Unlike the DUNE solution, our system allows tuning of the photosensor bias by adjusting the input laser power.
We also demonstrate the operation of the optical converter at temperatures down to 4.6~K, opening the possibility of using this technology in a much broader range of applications.
}

\keywords{Cryogenics; Lasers; Noble liquid detectors (scintillation, ionization, double-phase); Time Projection Chamber (TPC) }

\arxivnumber{2512.21207} % Only if you have one

\begin{document}
\maketitle
\flushbottom

\section{Introduction}
\label{sec::intro}

The Power over Fiber (PoF) technology transmits laser power over a non-conductive optical fiber to an Optical Power Converter (OPC) to power electrical devices or sensors. 
The main advantages of this technology are the removal of noise induced by standard power lines, voltage isolation, and spark-free operation,  thus offering robustness in a hostile environment~\cite{Articolo_PoF}.
It is therefore an ideal solution in environments where copper lines are prohibitive, such as for powering the Photon Detection System (PDS) in the DUNE Vertical Drift (VD) detector~\cite{DuneVD}.\\
DUNE is an experiment that will use liquid argon Time Projection Chambers (LAr TPC) as the detector to study neutrino oscillations, mass hierarchy, and CP violation in the leptonic sector~\cite{proposal_dune, dune_vol2, dune_vol1, dune_vol4}. 
A LAr TPC enables the reconstruction of the event topology and the estimation of the deposited energy by using the ionization-electron signal collected on the anode planes and the scintillation light collected by the PDS.
Furthermore, the PDS provides timing information and should ensure good photo-detection coverage.
The PDS of the DUNE VD relies on X-ARAPUCA devices, which are light traps for the LAr scintillation light, with SiPMs as photosensors~\cite{xarapuca}.
The X-ARAPUCAs will be placed on the high-voltage ($\sim$300 kV) cathode surface and immersed in liquid argon, in a prohibitive condition to safely operate with copper cables. For this reason, the ideal solution is to replace them with Power over Fiber~\cite{DuneVD}.

Existing PoF systems are commonly employed for voltage isolation between the source and the receiver, but they are not guaranteed to operate at cryogenic temperatures, such as those of liquid argon or liquid nitrogen~\cite{pof_matsuura,pof_rosolem}.
An extensive R$\&$D for the application of Power over Fiber at cryogenic temperature started at Fermi National Laboratory (FNAL, USA) in 2020 and demonstrated the feasibility of using it in the DUNE VD detector~\cite{pof_fnal}. One of the main drawbacks of this solution is its lack of tunability, which requires setting the photosensor bias to a fixed value at the start of the experiment.

The Cryo-PoF project was conceived in this framework, with the aims of powering both the photosensors (SiPMs) and their cold electronics using a single Power over Fiber line, while tuning the SiPM voltage bias as a function of the laser power. 
%Our method employs an operation region of the PoF combined with a custom DC-DC converter, where the SiPM bias voltage depends on the laser input bias voltage and, consequently, on the laser intensity. 
The novelty of our method lies in the use of PoF in combination with a custom DC-DC converter, with the SiPM voltage depending on the laser input control voltage and, consequently, on the laser intensity.
The Cryo-PoF line is composed of commercial products and employs a gallium arsenide (GaAs) laser source that transmits optical power through a multimode optical fiber to the optical power converter (OPC). 
The OPC converts the optical power into electrical power, with an efficiency of~$\sim$ 35\% at liquid nitrogen temperature, and it is capable of powering the photosensor electronic board.
Since the SiPMs require a higher bias voltage than the OPC output voltage, a DC-DC boost converter must be employed, whose output is tunable by slightly changing the laser power.
This feature allows fine-tuning of the SiPMs bias through the laser power, without the need for ancillary fibers.

The paper is organized as follows.
The laser transmitter and the optical fiber characterization are described in Section~\ref{sec::laser}. 
Section~\ref{sec::opc} describes the OPC performance both at room and cryogenic temperatures, with special attention to its operation at very low temperatures.
The description of the Cryo-PoF setup with the tunable DC-DC boost converter and the tests in liquid nitrogen to power the SiPMs are presented in Section~\ref{sec::testln}.

\section{Characterization of the laser source}
\label{sec::laser}

\begin{figure}[!htpb]
\centering
\includegraphics[keepaspectratio=true,scale=0.34]{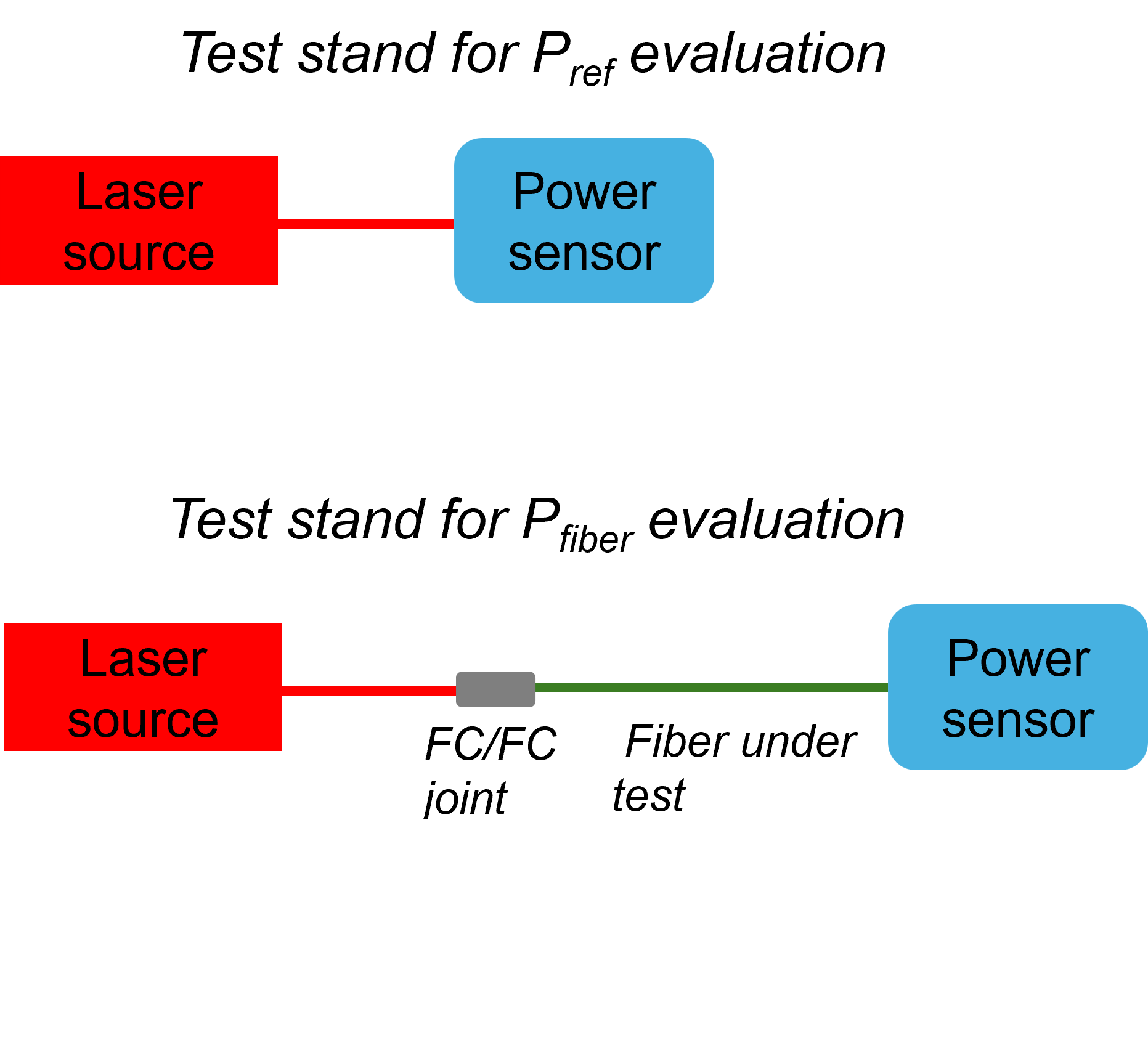}
\includegraphics[keepaspectratio=true,scale=0.34]{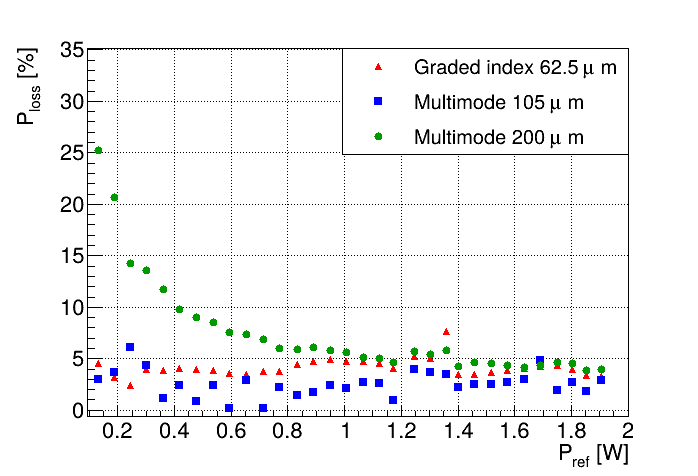} 
\caption{ Left up: sketch of the laser characterization test stand at room temperature to evaluate $P_{ref}$. Left bottom: sketch of the laser characterization test stand at room temperature to evaluate $P_{fiber}$. Right: power loss evaluated for each optical fiber tested using  eq.~\ref{eq::fiber}. The red triangles correspond to Fiber I, the blue squares to Fiber II and the green dots to Fiber III.} 
\label{fig::loss_con_fibre}
\end{figure}

The input source is an 808 nm gallium arsenide (GaAs) AFBR-POMEK2204 laser, manufactured by Broadcom~\cite{Laser_manuale}.
Its wavelength was chosen to be as far as possible from the SiPMs' photon detection efficiency peak (~$\sim$~400~nm~\cite{FBK_article}~), providing good power and efficiency for our purposes while reducing as much as possible any optical background to the photosensors. \\
The laser box has a 62.5 $\mu$m pigtail optical fiber and it is terminated with an FC/PC ceramic ferrule; the output power can be tuned by adjusting the input control voltage, reaching up to 2 W at room temperature.
We characterized the laser source at room temperature in terms of linearity between input control voltage and output power, power losses connecting an additional optical fiber, and stability over time. 
The laser characterization test stand, sketched in Figure~\ref{fig::loss_con_fibre}, includes the laser box, a connection optical fiber directly joint to the laser fiber through an FC/FC coupler, and a Thorlabs S142C Si sphere photodiode sensor read by a Thorlabs PM100D powermeter.
The additional fiber is necessary because the laser fiber is not long enough to reach the dewar, where the tests will be carried out.\\
To select such a fiber, we tested three multimode Thorlabs optical fibers to identify the one with the minimum power loss.
These fibers have a core, a cladding, a coating, and a reinforced 3.8 mm diameter black Furcation Tubing.
The Furcation Tubing consists of an outer PVC jacket, Kevlar$\textsuperscript{TM}$ protective threads, a polypropylene inner fiber tube, and a pull string for fiber insertion, which provides additional light blocking.
The reinforced tubing is necessary because, in preliminary tests, optical fibers without it showed significant light leakage, becoming an important background source for the SiPMs. 
The characteristics of the tested optical fibers are summarized in Table~\ref{tab:fibers}. \\
All tested fibers are 1.5~m long and terminate with an FC/PC ceramic ferrule connector. 
These optical fibers have a pure silica core and a fluorine-doped silica cladding.
\begin{table}[htpb]
    \centering
    \begin{tabular}{|c|c|c|c|}
    \hline
      Item   &  Fiber I & Fiber II & Fiber III \\
      \hline \hline
       Index Profile  & Graded index & Step-index  & Step-index \\
       \hline
       Numerical Aperture (NA)  & 0.27 & 0.22 & 0.22\\
       \hline
       Core $\varnothing$  &  62.5 $\mu$m & 105 $\mu$m & 200 $\mu$m\\
       \hline
       Cladding $\varnothing$  &  125 $\mu$m & 125 $\mu$m & 220 $\mu$m\\
       \hline
       Coating, $\varnothing$  &  Acrylate, $\varnothing$245 $\mu$m & Acrylate, $\varnothing$250 $\mu$m & Polyimide, $\varnothing$240 $\mu$m\\
       \hline
    \end{tabular}
    \caption{Characteristics of the Thorlabs optical fibers. The core material is pure silica and the cladding material is fluorine-doped silica for all tested fibers. They are 1.5 m long and have a $\varnothing$ 3.8~mm reinforced black tube. }
    \label{tab:fibers}
\end{table}
%\noindent

Each fiber was tested at room temperature by recording the output power as described above. The power loss $P_{loss}$ was estimated as:
\begin{equation}
    P_{loss} = \frac{P_{ref} - P_{fiber}} {P_{ref}} .
    \label{eq::fiber}
\end{equation}
%\noindent
where $P_{ref}$ is the power recorded at the end of the pigtail fiber (without any additional fiber) and $P_{fiber}$ is the power registered at the end of the tested fiber.
Considering the average $P_{loss}$, Fiber I shows a loss of $\sim$~4.44~$\pm$~0.41~\%, Fiber II of $\sim$~2.36~$\pm$~0.32~\%, and Fiber III of $\sim$~6.44~$\pm$~1.00~\%.
Therefore, the selected fiber was the multimode optical fiber with a 105 $\mu$m core diameter. 
All the results presented here were obtained with this fiber connected to the laser output.
Therefore, the reference power in all measurements is the power recorded at the end of the fiber.

\begin{figure}[htpb]
\centering
\includegraphics[keepaspectratio=true,scale=0.305]{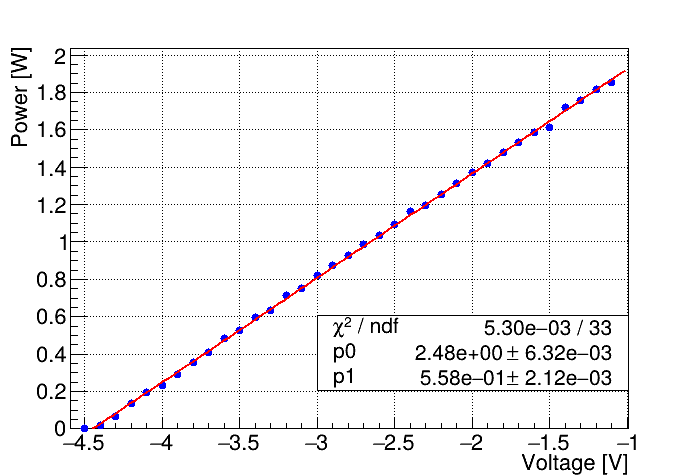}
\includegraphics[keepaspectratio=true,scale=0.305]{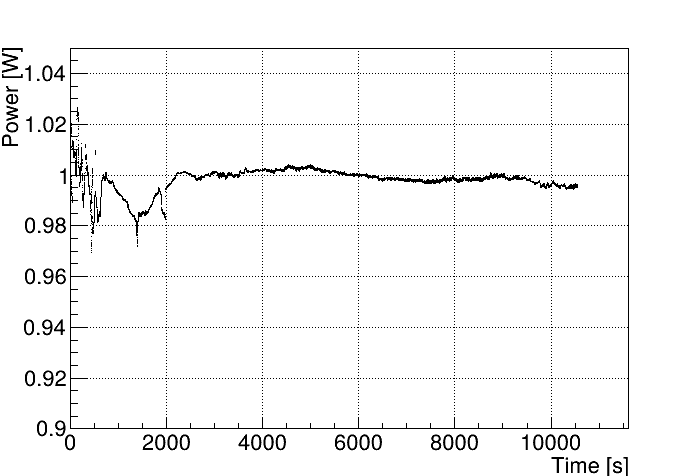}
\caption{Left: output laser power as a function of the input control voltage. Right: laser power stability at P$_{0}$~$\sim$~1~W.
\mbox{$(P_{max}-P_{min})/P_{0} \sim 5.7\%$}, with an average deviation from P$_{0}$ of 17.1~mW, where P$_i$ is the power measured at time $i$.
Excluding the first 30 minutes of operation, during which higher instability is observed, we obtain 
\mbox{$(P_{max}-P_{min})/P_{0} \sim 0.96\%$} and 
$\langle P_{0} - P_i \rangle = 15.9$~mW (see text for details)~\cite{cryo_pof_creta}.  } 
\label{fig::linearity}
\end{figure}

Since the laser source output is proportional to the input bias, we recorded the output power as a function of the input control voltage, in steps of 0.1 V, as plotted in Figure~\ref{fig::linearity}.\\
The plot confirms the linear behavior between the laser input control voltage and the delivered power.
The laser stability over time is a key parameter to bias the photosensors.
We continuously monitored the laser output power for two hours, setting an initial power of P$_{0}$~$\sim$~1~W (see Figure~\ref{fig::linearity} left)~\cite{cryo_pof_creta}.
Considering the difference between the maximum and the minimum values, we demonstrated good stability over time:
\mbox{$(P_{max}-P_{min})/P_{0} \sim 5.7\%$},
with an average deviation from $P_{0}$ of $\langle P_{0} - P_i \rangle = 17.1$~mW, where $P_i$ is the power measured at time $i$.
%Excluding the first 30 minutes of operation, during which higher instability due to thermalization  can be observed, excellent results of  
%\mbox{$(P_{max}-P_{min})/P_{0} \sim 0.96\%$} and 
%\mbox{Mean(P$_{0}$ – P$_i$) = 15.9~mW} were found.
During the first 30 minutes of operation, a higher instability was observed, mainly due to typical laser start-up effects, such temperature fluctuations and laser current instabilities \cite{pof_fnal}.
Excluding the initial period, excellent stability was achieved, with  
\mbox{$(P_{max}-P_{min})/P_{0} \sim 0.96\%$} and 
$\langle P_{0} - P_i \rangle = 15.9$~mW.

%\begin{figure}[htpb]
%\centering
%\includegraphics[keepaspectratio=true,scale=0.35]{Figure/Stabilita_1W_grande.png}
%\caption{Laser power stability at fixed input power of P$_{0}$~$\sim$~1~W.
%\mbox{$(P_{max}-P_{min})/P_{0} \sim 5.7\%$}, with an average deviation from P$_{0}$ of  = 17.1 mW, where P$_i$ is the power measured at time $i$.
%Excluding the first 30 minutes of operation, during which higher instability is observed, we obtain 
%\mbox{$(P_{max}-P_{min})/P_{0} \sim 0.96\%$} and 
%$\langle P_{0} - P_i \rangle = 15.9$~mW (see text for details)~\cite{cryo_pof_creta}.}
%\label{fig::stabilita}
%\end{figure}

\section{The Optical Power Converter}
\label{sec::opc}

\begin{figure}[htpb]
\centering
\includegraphics[keepaspectratio=true,scale=0.305]{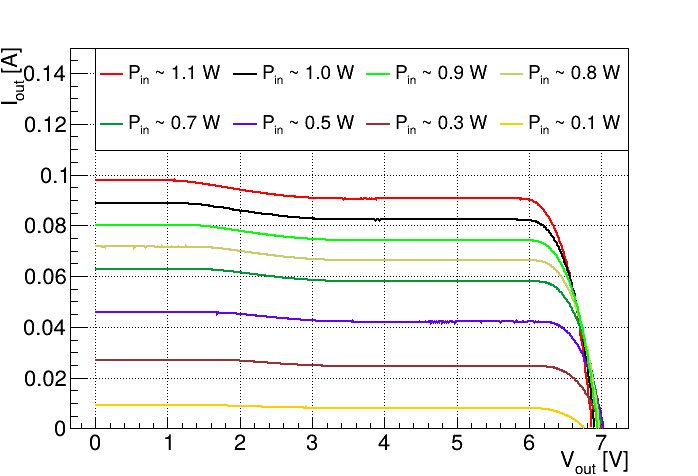}
\includegraphics[keepaspectratio=true,scale=0.305]{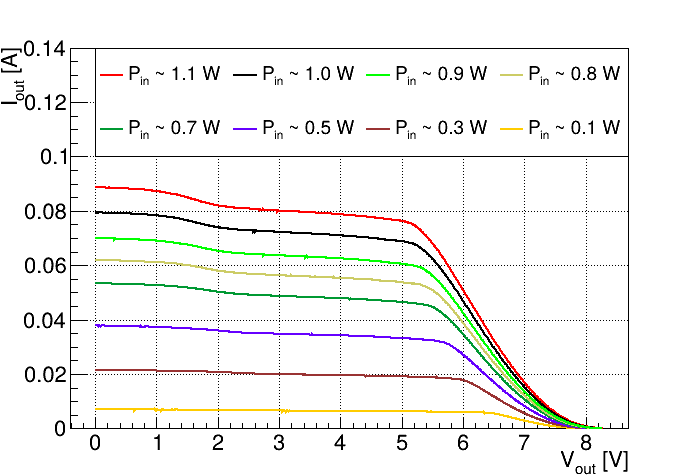}
\caption{IV curves for the OPC at room (left) and liquid nitrogen temperature (77 K -- right).} 
%for all the laser powers delivered during the tests.}
\label{fig::IV_OPC}
\end{figure}

\begin{figure}[htpb]
\centering
\includegraphics[keepaspectratio=true,scale=0.305]{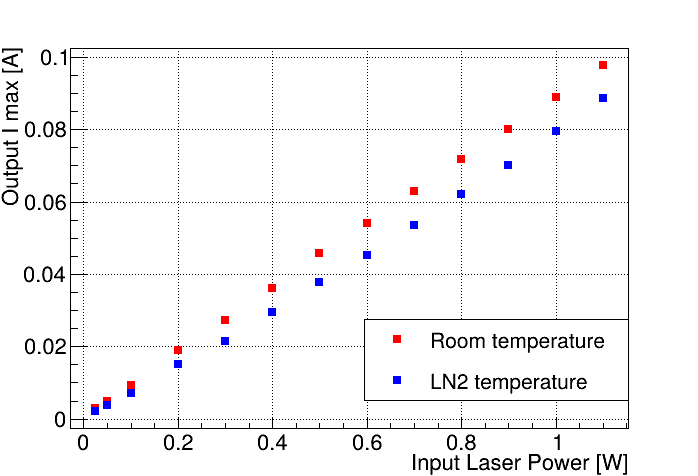}
\includegraphics[keepaspectratio=true,scale=0.305]{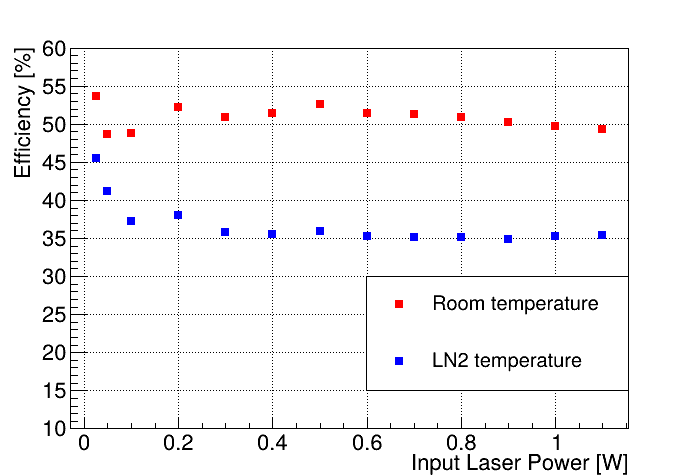}
\caption{Maximum current delivered by the OPC at different laser powers (left) and OPC efficiency as a function of the laser power (right). Error bars are included but hidden by the data points. In each plot, the red dots correspond to measurements at room temperature, while blue dots correspond to measurements at 77 K.}
\label{fig::eff_opc}
\end{figure}

The key component of the PoF line is the Optical Power Converter (OPC), manufactured by Broadcom~\cite{opc_manuale}, that transforms the laser light into a DC voltage up to 7~V. 
The optical fiber is directly inserted in the FC connector of the OPC, and its output voltage can be changed by acting on the input laser power. 
The OPC characterization was performed by evaluating the maximum power and current at both room and cryogenic temperatures (in a liquid nitrogen bath at 77 K) for different $P_{in}$ powers.  
The temperature was monitored by a PT100 placed near the device to ensure that it reached cryogenic conditions.
We measured the current-voltage (IV) curve by means of a Keithley$\textsuperscript{TM}$ 4200A-SCS semiconductor parameter analyzer. 
In Figure~\ref{fig::IV_OPC}, we report the IV curves for a wide range of P$_{in}$, both at room and cryogenic temperatures.
%noindent
From the IV curves, it is possible to obtain information on both the maximum current delivered $I_{max}$ and the output power, and thus on efficiency.
The maximum current increases linearly with the laser power, as reported in Figure~\ref{fig::eff_opc}.

\begin{figure}[htpb]
\centering
\includegraphics[keepaspectratio=true,scale=0.30]{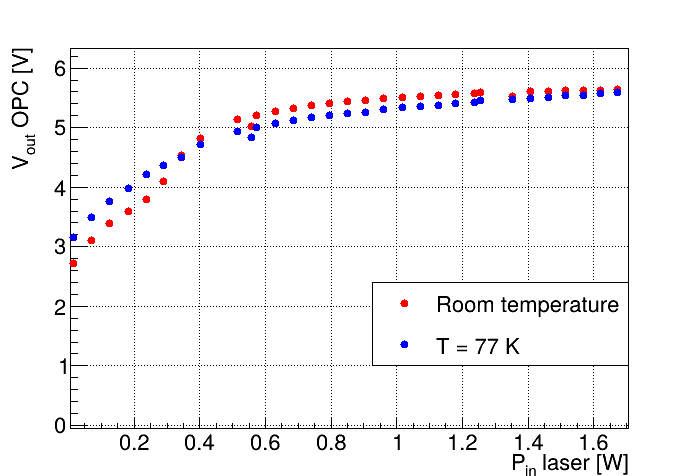}
\includegraphics[keepaspectratio=true,scale=0.30]{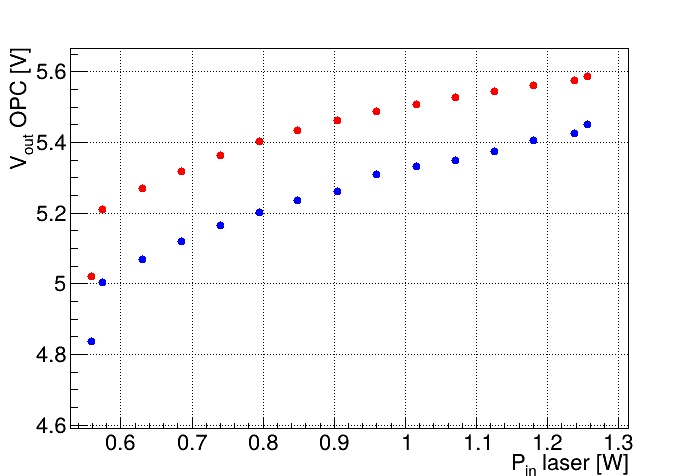}
\caption{OPC output voltage $V_{out}$ as a function of the input laser power $P_{in}$. $V_{out}$ corresponds to the input voltages for the DC-DC boost converter and the SiPM amplifier.  Error bars are included but hidden by the data points. The bias tuning of the SiPMs was achieved by exploiting the region between 0.7 W and 1.2 W, as displayed in the right plot. In both plots, red points correspond to measurements taken at room temperature and blue points to those taken at cryogenic temperatures.}
\label{fig::Vout_opc}
\end{figure}

The output power of the OPC ($P^{OPC}$) is obtained by the IV curves and amounts to:
\begin{equation}
    P^{OPC} = I \cdot V \ .
\end{equation}
%\noindent
The OPC maximum efficiency \textit{Eff} is given by:
\begin{equation}
    \textit{Eff} = \frac{P_{max}^{OPC}} {P_{in}},
\end{equation}
%\noindent
where $P_{max}^{OPC}$ is the maximum output power provided by the OPC and  $P_{in}$ is the power delivered by the laser source.
The IV curves also provide information on the maximum current delivered by the OPC.
For example, considering P$_{in}$ = 1 W, the efficiency at room temperature is \textit{Eff}$\simeq$50\%, with a maximum delivered current of $I_{max}$~$\simeq$~88.9~mA, while the efficiency decreases down to \textit{Eff}$\simeq$35\%, with an I$_{max}$~$\simeq$~79.5~mA, at liquid nitrogen temperature.\\
In Figure~\ref{fig::eff_opc}, the efficiency and maximum current delivered by the OPC for each tested P$_{in}$ are reported. The efficiency strongly depends on the laser power for low $P_{in}$, both at room and cryogenic temperature, while for $P_{in}$ $>$ 0.4 W, it reaches a stable value of $\sim$50\% at room temperature and $\sim$35\% at cryogenic temperature. 
%The maximum current increases linearly with the laser power, as reported in Figure \ref{fig::eff_opc}. 
%
%\noindent
The output voltage of the OPC was measured with a load, i.e. connected to SiPMs and electronics, to mimic real experimental conditions.
The maximum voltage obtainable from the OPC under these conditions, around 5 V, varies only slightly with the laser power, both at room temperature and in a liquid nitrogen bath (see Figure \ref{fig::Vout_opc}). 
The region between 0.7 W and 1.2 W results the ideal operating range for the Cryo-PoF tunable system and it is suitable for stably powering all components using a single OPC and laser source, as highlighted in Figure \ref{fig::Vout_opc} right.

\begin{figure}[htpb]
\centering
\includegraphics[keepaspectratio=true,scale=0.4]{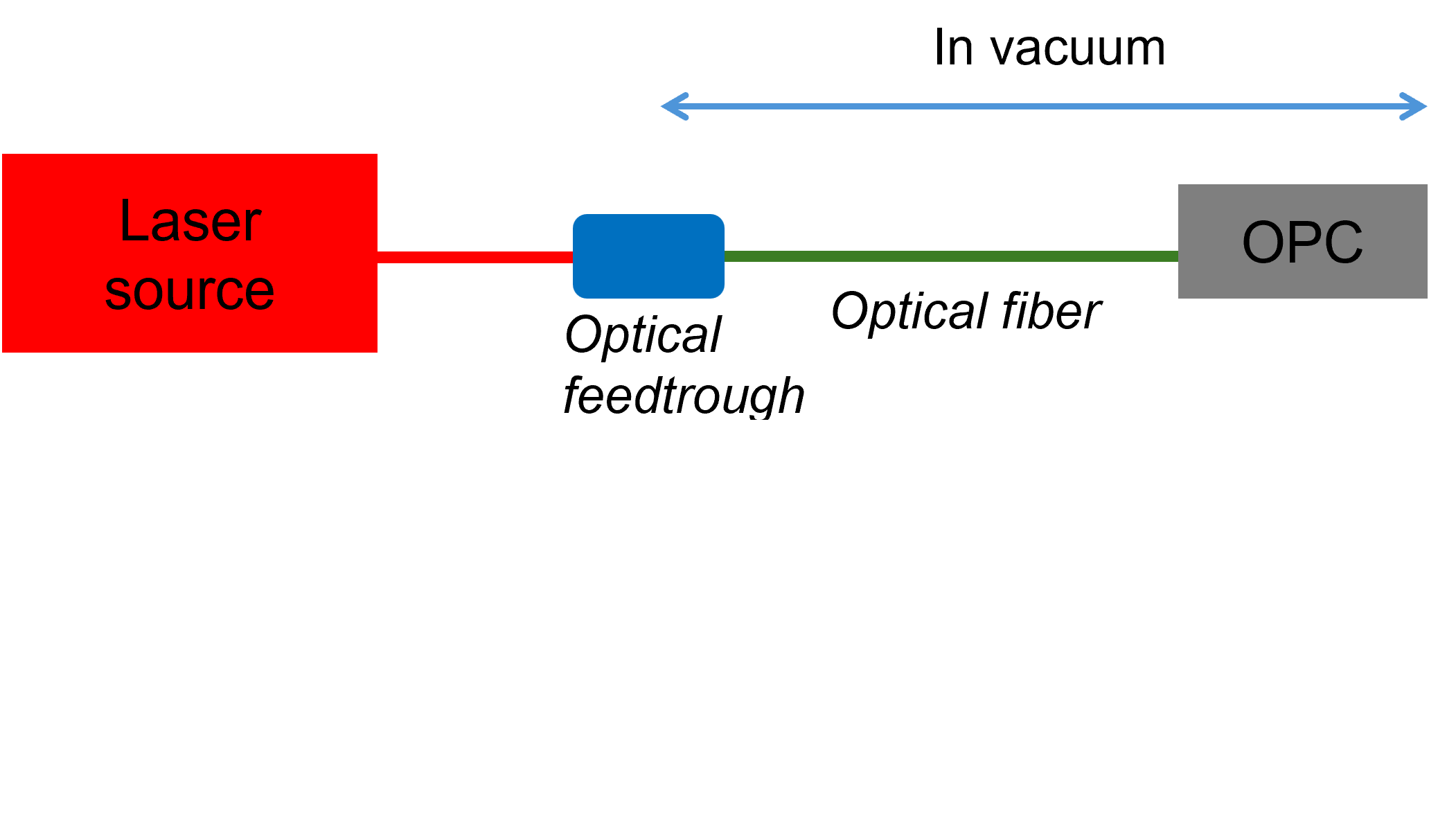}
\hspace{1 cm}
\includegraphics[keepaspectratio=true,scale=0.8]{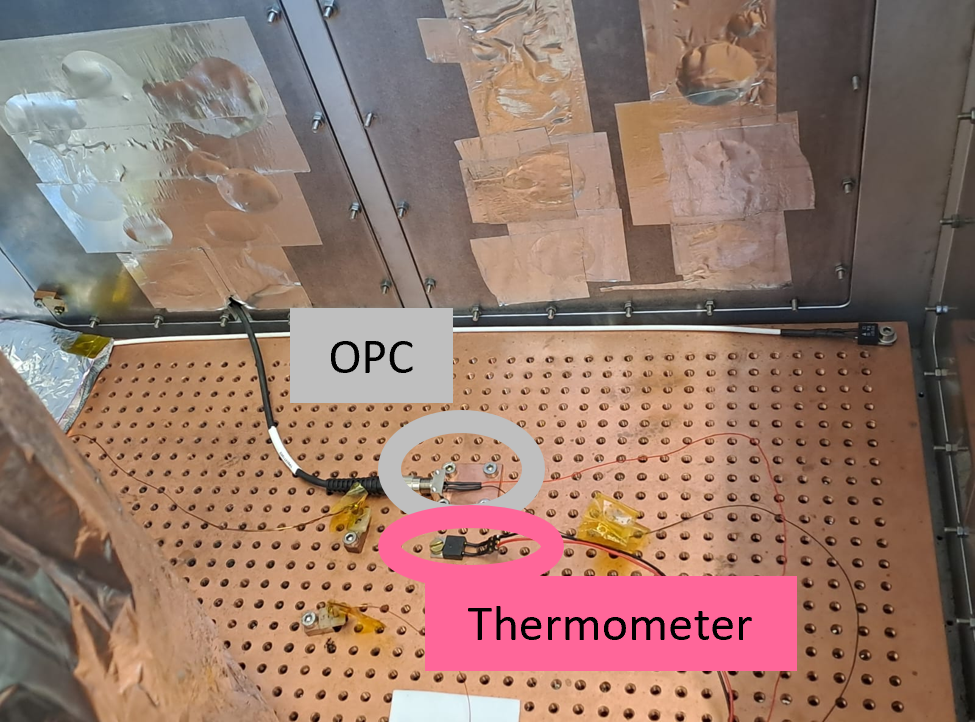}
\caption{Setup of the measurements at very low temperatures (left) and inside the cryostat (right). The OPC is in contact with the plate, as is the thermometer, for precise temperature control. }
\label{fig::low_temp_setup}
\end{figure}

\begin{figure}[htpb]
\centering
\includegraphics[keepaspectratio=true,scale=0.34]{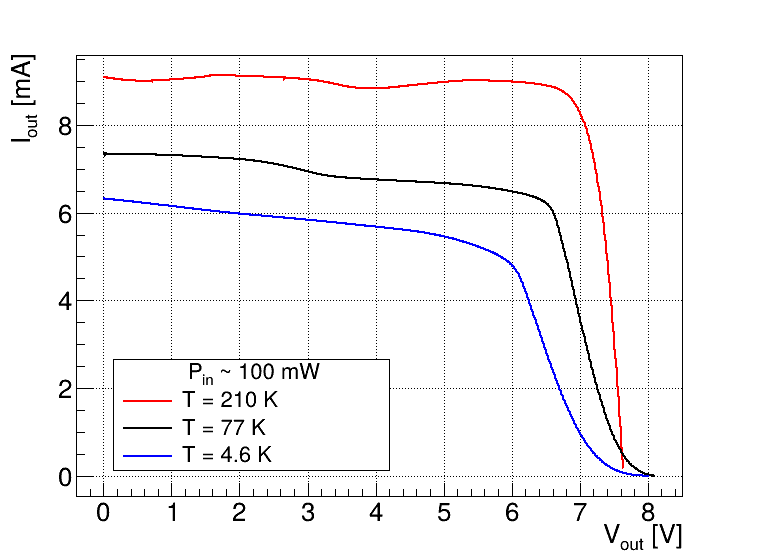}
\caption{IV curves of the OPC in the cryostat at low temperature. The OPC reliably operates down to 4.6 K, with an efficiency of 29\%.}
\label{fig::lowT}
\end{figure}

In view of applications beyond high-energy physics, we tested the OPC in a cryostat capable of reaching temperatures down to few K degrees.
Preliminary results, obtained with a non-optimised light transmission chain, are published in~\cite{cryo_pof_elba}.\\
The 808~nm GaAs laser source optical fiber (62.5~$\mu$m core) was connected to an optical feedthrough with a 100~$\mu$m core diameter, which permits entry in the cryostat, as sketched in Figure~\ref{fig::low_temp_setup} left. 
Inside the cryostat, we used the same 105~$\mu$m multimode optical fiber as in the liquid nitrogen measurements, equipped with a vacuum-compatible termination and connected to the OPC. 
The OPC was mounted on the cryostat copper plate, near a thermometer for temperature monitoring (see Figure \ref{fig::low_temp_setup} right).
The characterization of the OPC output was performed by recording the IV curves through the same Keithley$\textsuperscript{TM}$ 4200A-SCS semiconductor parameter analyzer.
Before closing the cryostat, we set and validated the power at the OPC with the powermeter, in order to have P$_{in} \simeq$ 100 mW inside the cryostat. As a check, we recorded the IV curves with the cryostat under vacuum at room temperature, which were found to be identical to those obtained at atmospheric pressure.
As can be seen in Figure~\ref{fig::lowT}, the OPC works down to the minimum temperature that can be achieved by the cryostat -- 4.6 K --  with an efficiency of 29\%. This efficiency is only marginally lower than the one obtained in liquid nitrogen.

\section{Test with photosensors and a tunable DC-DC boost converter}
\label{sec::testln}

\begin{figure}[htpb]
\centering
\includegraphics[keepaspectratio=true,scale=0.40]{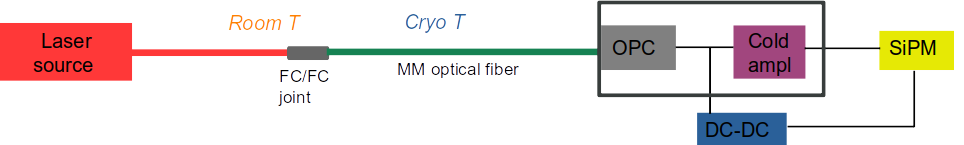}
\caption{Schematic representation of the Cryo-PoF setup. The laser light power is converted into electrical power by the OPC, which supplies the DC-DC boost converter, the SiPMs, and the cold amplifier. The laser source operates at room temperature, while the other components are maintained at cryogenic temperatures. }
\label{fig::schema}
\end{figure}

A schematic of the Cryo-PoF setup used to power the photosensors is shown in Figure \ref{fig::schema}.
It is composed of the laser source, connected via the optical fiber described in Section~\ref{sec::laser} to the OPC described in Section~\ref{sec::opc}, the electronic board~\cite{DuneVD}, and the photosensors.
The electronic board houses the OPC, together with the cold amplifier and the DC-DC boost converter.\\
The 20 SiPMs, connected together in hybrid ganging mode, are mounted on the same mechanical support. 
They share the same input/output cable and constitute a flex (see Figure~\ref{fig::cold_amp} for the SiPM ganging scheme in a flex and its connection to the I/O cable).
Four flexes, i.e. 80 SIPMs, represent a DUNE acquisition channel and are biased at the same voltage.
The DUNE VD detector will be equipped with two different custom SiPM models:  Hamamatsu (HPK) S13360-9935-75 \textmu m-HQR~\cite{HPK_article}  and FBK NUV-HD-Cryo 3T~\cite{FBK_article}.
They have different breakdown voltages (V$_{bd}^{HPK} \simeq 42.0$~V and V$_{bd}^{FBK} \simeq 27.1$~V) and, therefore, two different DC-DC boost converter boards were used.

\subsection{The cold amplifier}
\label{subsec::coldampl}

\begin{figure}[htpb]
\centering
\includegraphics[keepaspectratio=true,scale=0.60]{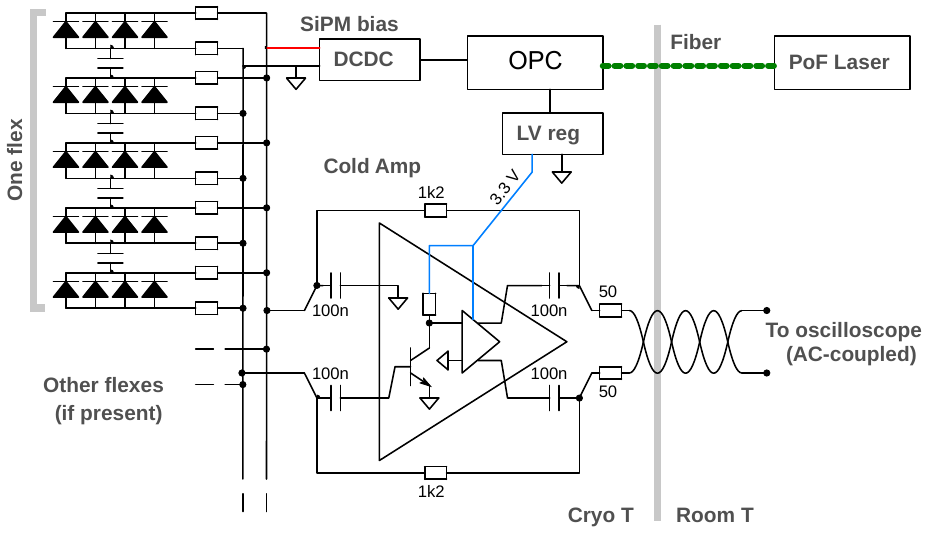}
\caption{Circuit diagram of the cold amplifier. The connection to the SiPMs, the laser and the oscilloscope are also highlighted.}
\label{fig::cold_amp}
\end{figure}

The cold amplifier has been developed by the DUNE Milano Bicocca group~\cite{ampl_article}. It is an inverting transimpedance amplifier, which converts the current signals from the SiPMs to differential voltage signals to be recorded by the signal acquisition chain. 
The circuit is based on a SiGe heterojunction bipolar input transistor (BFP640), followed by a fully differential opamp (THS4531). It draws $0.7\ \textrm{mA}$ from a single $3.3\ \textrm{V}$ supply; it is powered directly by the OPC, through a level regulator. The series white noise at 77 K is $0.4\ \textrm{nV}/\sqrt{\textrm{Hz}}$.
The closed-loop bandwidth is  $5\ \textrm{MHz}$, which gives a signal rise time of about $70\ \textrm{ns}$.
The signal fall time is given by the recovery time of the SiPM cells. The amplification chain we used in all our measurements provides a 20x gain. 
The design of the cold amplifier is described in detail in~\cite{ampl_article} and schematized in Figure~\ref{fig::cold_amp}. 
Its performance in reading out different numbers of SiPMs in various ganging configurations are described in~\cite{ampl_article2}.

\subsection{The DC-DC boost converter}
\label{subsec::dcdc}

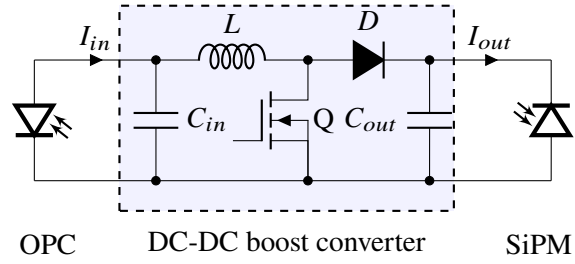
\begin{figure}[htpb]
\centering
\begin{circuitikz}[american, scale=0.8]
\ctikzset{
    resistors/scale=0.7,
    capacitors/scale=0.7,
    diodes/scale=0.7,
}
% PV module label
\fill[fill=white!5] (-.5,-1.5) rectangle (0.95,1.8)
node[midway,below=1.5cm,black]{{OPC}};

% Boost converter label
\fill [blue!5] (1.4,-1.5) rectangle (6.9,1.9)
    node[midway,below=1.5cm,black]{{DC-DC boost converter}};

% Linea tratteggiata intorno al blocco Boost Converter
\draw[dashed, thick, black] (1.4,-1.5) rectangle (6.9,1.9);

% Load label
\fill [fill=white!5] (7.6,-1.5) rectangle (9,1.8)
    node[midway,below=1.5cm,black]{SiPM};

\draw (0,1) to [short,i=$I_{in}$] ++(2,0) coordinate(a1) node[circ]{}
    to[cute inductor,l=$L$] ++(2.5,0) coordinate(b1) node[circ]{}
    to[D*,l=$D$] ++(2,0)
    coordinate(c1) node[circ]{}
    to [short,i=$I_{out}$] ++(2,0)coordinate(d1);
 
% Bottom horizontal path (L,D) 
\draw (0,-1) -- ++(2,0) coordinate(a2) node[circ]{}
    -- ++(2.5,0) coordinate(b2) node[circ]{}
    -- ++(2,0) coordinate(c2) node[circ]{}
    -- ++(2,0) coordinate(d2);
 
% Add a resistor
\draw (d2) to[ empty photodiode, thick, -] (d1);
 
% Add input capacitor
\draw (a1) to[C,l=$C_{in}$] (a2);

\draw(0,1) to[empty photodiode, thick, -] (0,-1);
 
% Add output capacitor
\draw (c1) to[C,l_=$C_{out}$] (c2);
 
% Add switch 
\node[nigfete] (switch) at ($0.5*(b1)+0.5*(b2)$){ Q} ;
\draw (b1) node[]{}-- (switch.D);
\draw (b2) node[]{} -- (switch.E);
\end{circuitikz}
\caption{Schematic of the HPK DC-DC boost converter circuit.}
\label{fig:dc_dc}
\end{figure}

The DC-DC boost converter in use has been developed by the INFN Milano group; it raises the OPC output voltage to the level required for SiPM biasing. Typically, the OPC provides an output voltage that is several tens of volts lower than what is needed to properly bias the light sensors. For this reason, the DC-DC converter is expected to receive an input voltage V$_{in}$ of $~\sim$~5~V, and the voltage transfer function is determined by the duty cycle of the Pulse Width Modulation (PWM) control signal. Consequently, the control circuitry plays a crucial role in maintaining a stable output voltage, preventing the control loop from drifting and causing unintended voltage drops. A complete description and characterization of the system can be found in~\cite{dcdc_article} and~\cite{dcdc_article_25}, while a schematic of the circuit is shown in Figure~\ref{fig:dc_dc}.

%VT: se pensi che questa immagine non sia necessaria puoi cancellarla !!
%MT: va benissimo

%
%\noindent
The circuit is a non-synchronous boost converter. The inductor \textbf{L} stores energy when the power switch \textbf{Q} (MOSFET) is on and releases it to the output when \textbf{Q} is off through the rectifying diode \textbf{D}, thus producing $V_{\mathrm{out}} > V_{\mathrm{in}}$ (ideal relation: $V_{\mathrm{out}} \approx V_{\mathrm{in}}/(1-D)$, with duty cycle $D$). The input capacitor $C_{in}$ filters the current drawn from the OPC, while the output capacitor $C_{out}$ smooths the boosted voltage supplied to the SiPMs. A PWM controller with analog feedback (error amplifier, compensation network, and resistive divider) regulates the duty cycle to set the desired bias and reject perturbations; the setpoint is chosen according to the SiPM type and can be finely tuned by varying the laser power.

\begin{figure}[htbp]
\centering
\includegraphics[keepaspectratio=true,scale=0.305]{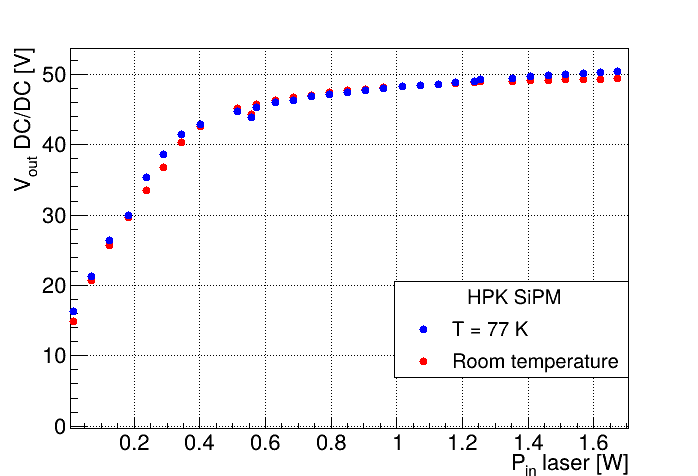}
\includegraphics[keepaspectratio=true,scale=0.305]{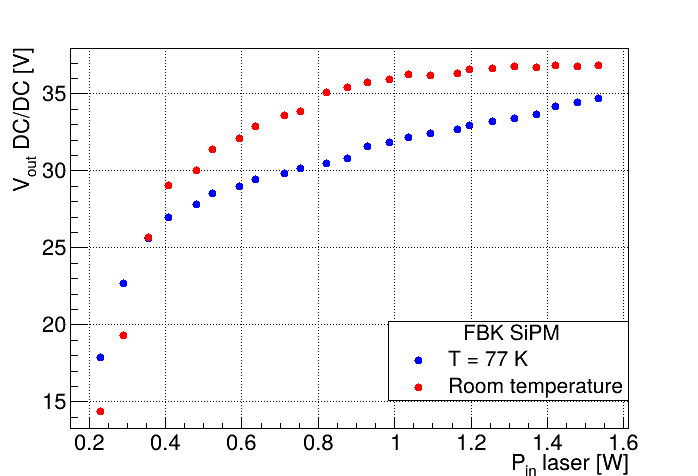}
\caption{DC-DC boost converter output voltage V$_{out}$ as a function of the input laser power, for the HPK (left) and FBK (right) SiPMs~\cite{cryo_pof_creta, cryo_pof_elba}.  The blue dots correspond to measurements in liquid nitrogen, while the red dots correspond to measurements at room temperature.} %The FBK DC-DC temperature dependence is due to the presence of a trimmer.}
\label{fig::DCDC}
\end{figure}

Since two different breakdown voltages $V_{bd}$ are expected due to the use of two different SiPM types, two separate DC–DC converter boards were employed, each tuned to the corresponding SiPM type. Specifically, they provide output voltages in the range of 40–50~V for the Hamamatsu SiPMs and 25–35~V for the FBK SiPMs, thus supplying the appropriate bias to each photosensor type~\cite{cryo_pof_creta,cryo_pof_elba}. \\
The V$_{out}$ vs V$_{in}$ dependency is essentially unaffected by temperature in the HPK DC-DC, while there is a difference for the FBK one (see Figure~\ref{fig::DCDC}). 
This behavior can be explained by a combination of effects. In particular, in the FBK board the output setpoint is adjustable through a sealed cermet trimmer placed in the feedback divider, which sets the PWM controller reference gain, i.e. $V_{\text{out}} = V_{\text{ref}}\!\left(1 + \frac{R_\mathrm{top}}{R_\mathrm{bot}}\right)$.

A TCR of the divider elements causes a small drift of the ratio $R_\mathrm{top}/R_\mathrm{bot}$ with temperature. We used a sealed cermet $100~\mathrm{k}\Omega$ trimmer from VISHAY (part number T36YB3), with a TCR value of $\pm 100$~ppm/$^\circ$C. Such a TCR on a 35~V bias corresponds to an equivalent drift of about 5~mV/$^\circ$C at the output, if the ratio is dominated by the adjustable leg. In our configuration, the HPK board uses only a fixed divider, whereas the FBK board includes the trimmer to accommodate the different breakdown voltage; this intentional design difference can explain a part of the temperature dependence observed in Figure~\ref{fig::DCDC}, accounting for approximately 1~V of the total room-to-LN$_2$ shift. Quantitatively, the FBK converter shows a room-to-LN$_2$ offset of approximately 5~V across the 25--35~V window. Two experimental aspects help explain the residual difference: the two boards were characterized in different output-voltage windows, hence at different control-loop operating points; moreover, the FBK unit uses a slightly different DC--DC prototype, with passive sizing and compensation optimized for the lower-voltage range. At cryogenic temperature, small variations in reference drift, error-amplifier input offsets and bias currents, as well as passive-component effects (ESR, dielectric behavior), can reduce the effective DC loop gain and produce a static set-point shift.
The $V_{\mathrm{out}}$--$V_{\mathrm{in}}$ curves in Figure~\ref{fig::DCDC} were taken on a minimal bench: the laser power was stepped within the 0.7--1.2~W window to sweep the OPC and set $V_{\mathrm{in}} \simeq 5$~V at the converter input, while the DC--DC output node was probed by a high-impedance external monitor and read with the Agilent U1232A multimeter; non-essential blocks (e.g., SiPMs and front-end) were not used in this specific test.
This ability to allow external adjustment of V$_{out}$ is a key feature of these DC–DC converters; the converters can set V$_{out}$ as a function of the laser power, enabling fine regulation of the photosensor bias simply by adjusting the laser power, without requiring additional fiber lines.  This feature is particularly valuable in experiments where the power units are inaccessible, such as when they are immersed in liquid argon or other cryogenic media. 

%AF Frasi ri-aggiustate per evidenziare solo l'uso del DC-DC in CryoPoF
%A key feature of these DC–DC converters is their ability to allow external adjustment of V$_{out}$, enabled by an external control signal whose duty cycle regulates the converter’s output voltage. In the absence of the remote control signal, or in the event of a failure, the converter automatically reverts to operating with the default output value. 
%The ability to select a specific output voltage greatly enhances the flexibility of the DC-DC converter, which is particularly valuable in experiments where the power units are inaccessible, such as when they are immersed in liquid argon or other cryogenic media.
%Another feature of the DC–DC converters has also proven useful in the current Cryo-PoF project, where the board has been employed in a configuration different from the one described above. In this setup, the converters can set V$_{out}$ as a function of the laser power (reported in Figure~\ref{fig::DCDC}), enabling fine regulation of the photosensor bias simply by adjusting the laser power, without requiring additional fiber lines.

\subsection{Test results} 
\label{subsec::test}

\begin{figure}[htbp]
\centering
\includegraphics[keepaspectratio=true,scale=0.45]{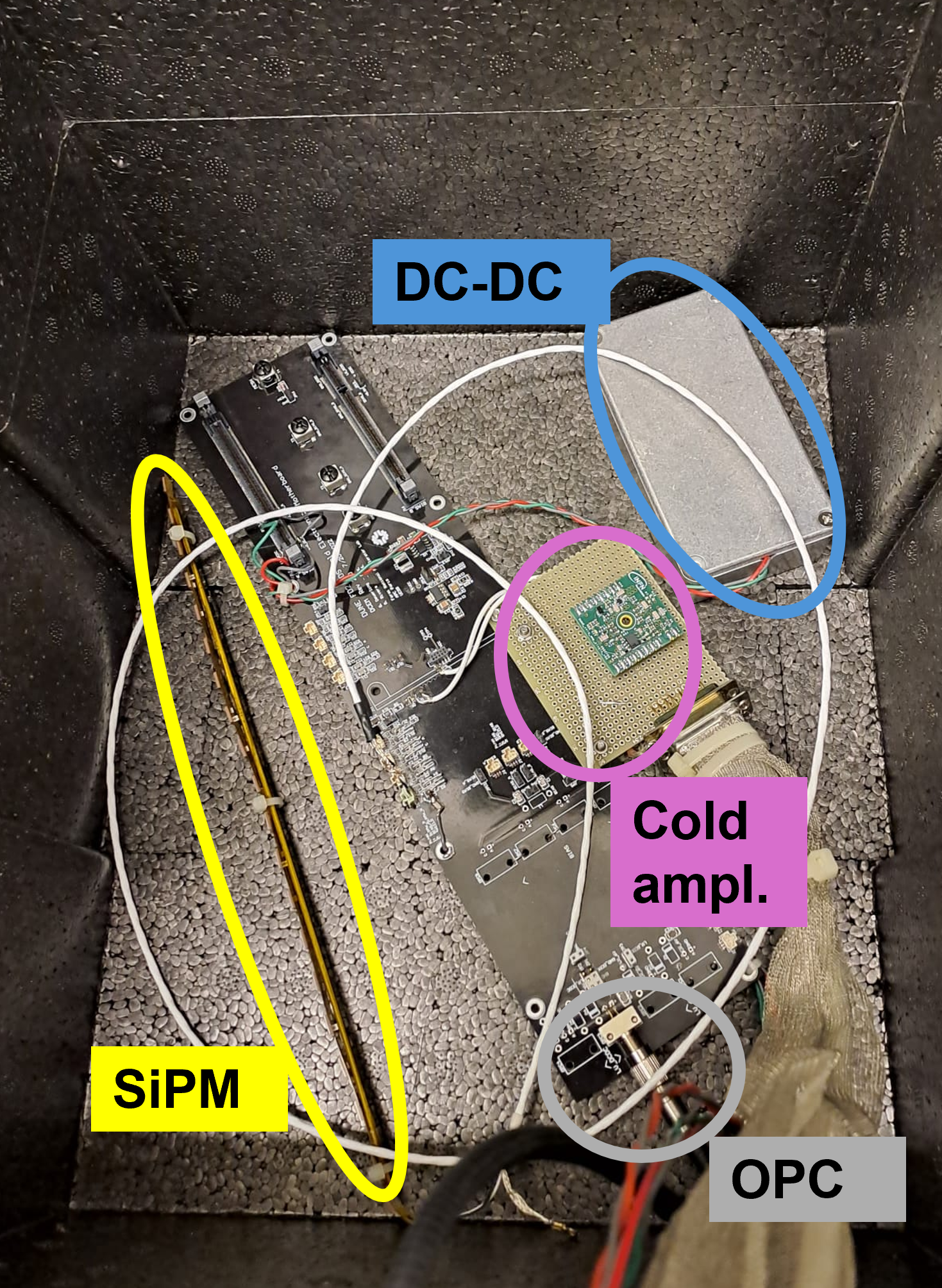}
\caption{Picture of the Cryo-PoF electronic board and setup used to power 20 Hamamatsu SiPMs: the OPC (gray), the amplifier (purple), the DC-DC converter inside the metallic box (blue) and the 20 SiPMs (yellow) are highlighted.}
\label{fig::scatola}
\end{figure}

In order to evaluate the Cryo-PoF performance, we powered 20 Hamamatsu SiPMs in a liquid nitrogen bath (see Figure~\ref{fig::scatola}), both with standard copper cables and with the Cryo-PoF system.
%\noindent
During the test, an induced noise from the DC-DC converter was observed and subsequently mitigated by placing the DC-DC board inside a metallic enclosure acting as a Faraday cage.
In our case, the adoption of a compact shield has negligible impact on the system envelope or mass and does not constrain the architecture, allowing the converter to be integrated on the main board without issues.

The SiPMs were illuminated by a 400 nm LED light driven by a SP5601 CAEN LED driver, which provides tunable 400 nm light and was used as the trigger source for the oscilloscope.
The amount of light was adjusted to have one or two photons detected by the SiPMs.\\
The tests were performed at three different bias voltages: 45 V, 46 V, and 47 V. 
The DC-DC output voltage directly biases the SiPMs. It was monitored by an external circuit, read with a multimeter and tuned by means of the laser power.
Once the bias value is set, 5$\times 10^3$ waveforms were recorded by the oscilloscope for each bias voltage.
\begin{figure} [htpb]
\centering
\includegraphics[keepaspectratio=true,scale=0.265]{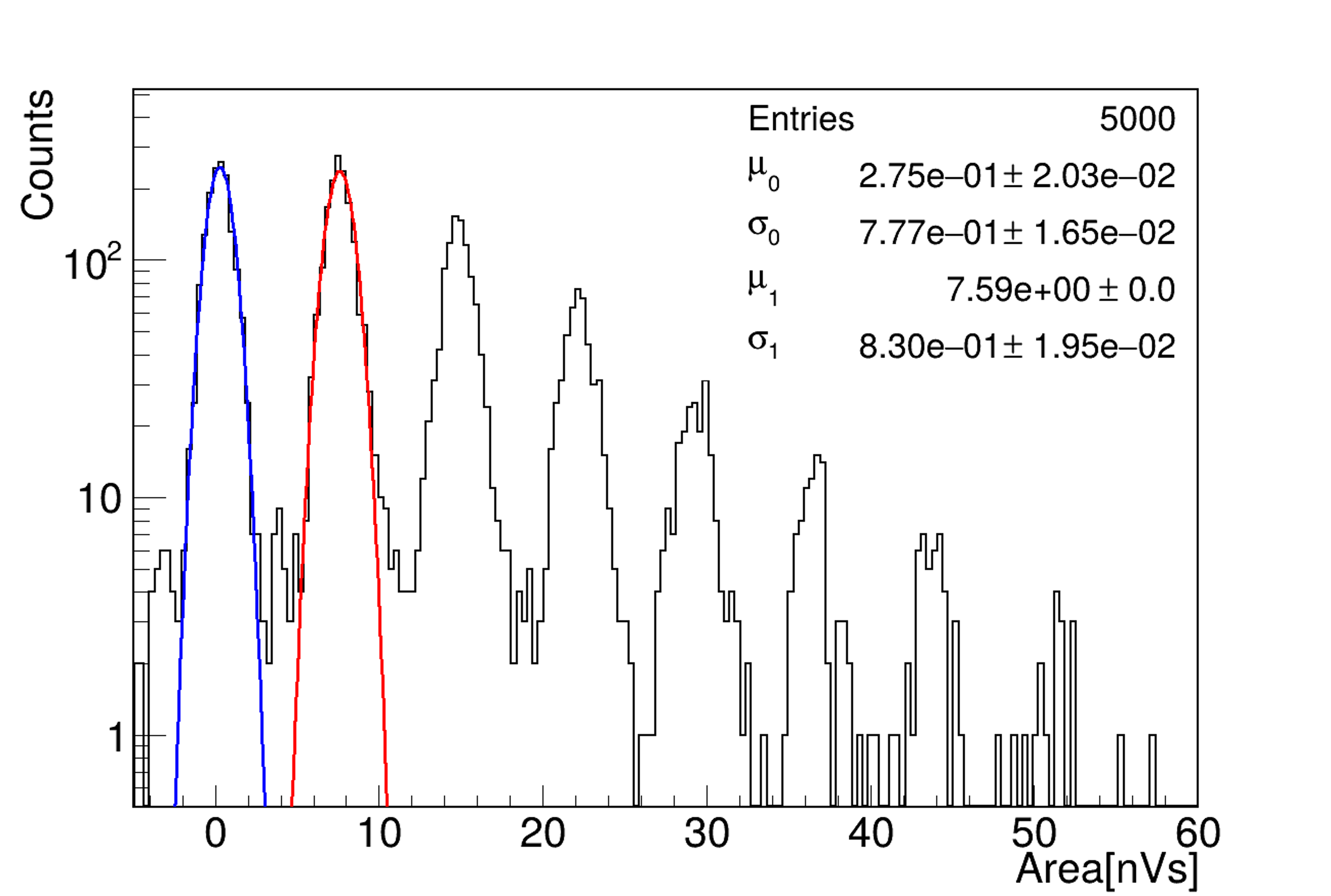}
\includegraphics[keepaspectratio=true,scale=0.265]
{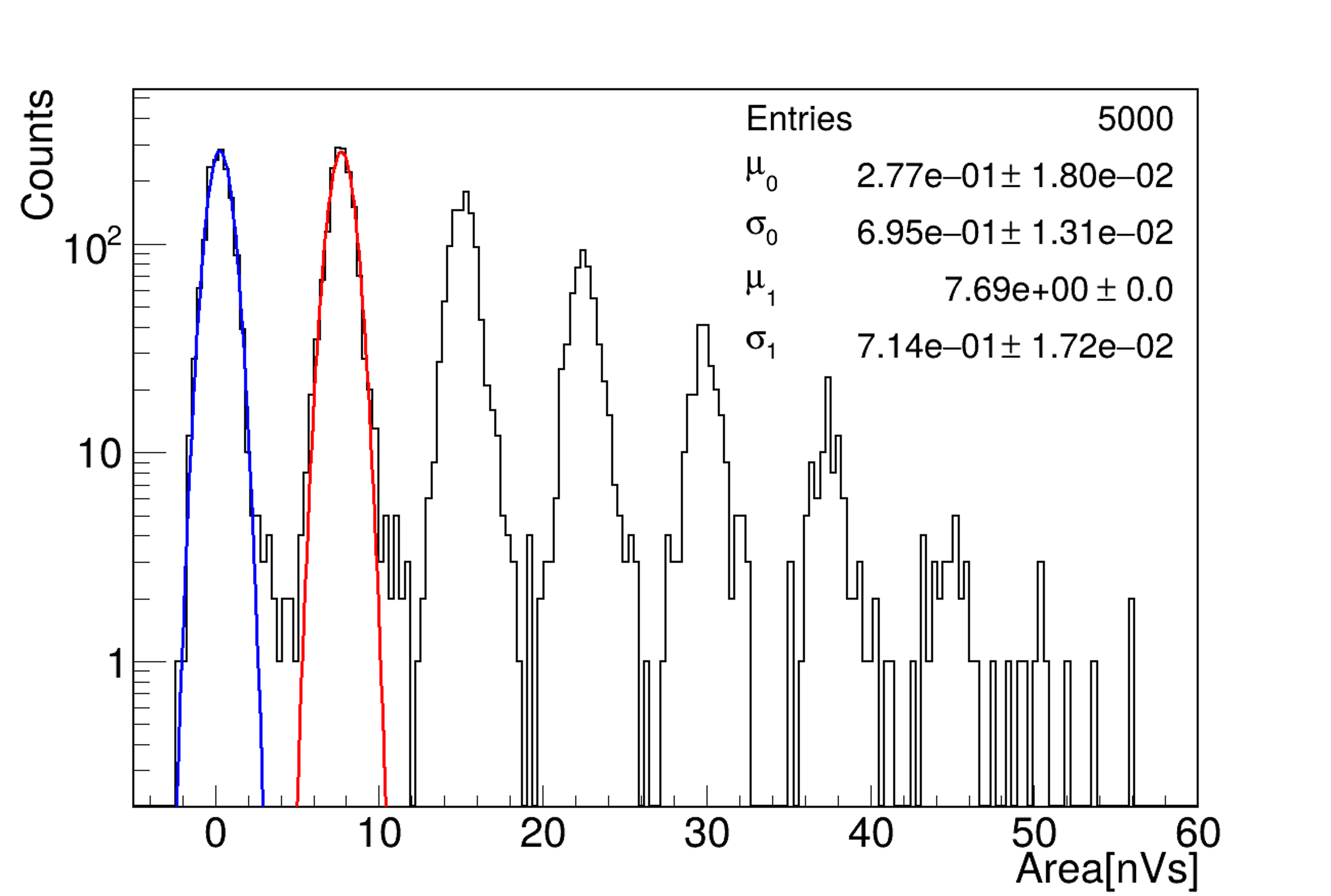}
\caption{Photoelectron spectra obtained from 20 SiPMs powered via Cryo-PoF (left) and a standard copper cable (right). Measurements were taken at V = 46 V, corresponding to a laser power of 1 W in the Cryo-PoF configuration. The SiPMs were illuminated with LED light, adjusted to yield approximately one or two detected photons per SiPM. Each histogram shows the integrals (time window of 900 ns) of 5$\times 10^3$ waveforms. The plots are fitted with a linear combination of two Gaussians, corresponding to 0 (blue curve) and  1 (red curve) recorded photoelectrons\cite{cryo_pof_creta}.}
\label{fig::snr}
\end{figure}
Figure~\ref{fig::snr} shows the photoelectron spectra obtained with the two configurations. 
Each waveform was integrated over a 900 ns time window, and the resulting integrals were fitted with a multi-Gaussian function.\\
To quantify the performance of the Cryo-PoF setup, we use the Signal-to-Noise Ratio (SNR), calculated as :
\begin{equation}
    SNR = \frac{\mu_1 - \mu_0}{\sigma_0},
\end{equation}
%
%\noindent
where $\mu_{1}-\mu_{0}$ is the gain of the SiPM and $\sigma_{0}$ is the width of the Gaussian corresponding to the noise peak. 
%As shown in Table~\ref{tab::risultati}~\cite{cryo_pof_creta}, the performance of the PoF system is comparable to that of the copper-cable setup. 
%Residual noise from the DC-DC converter is likely responsible for a slight degradation of the Cryo-PoF performance.
%

\begin{table} [htpb]
\begin{center}
 \resizebox{\textwidth}{!}{
    \begin{tabular}{|c|c|c|c|c|c|c|c|}
    \hline
   \multicolumn{1}{|c|}{P laser} &\multicolumn{1}{|c|}{SiPM bias} & \multicolumn{2}{|c|}{$\mu_1 - \mu_0$ (nV$\cdot$s)} & \multicolumn{2}{|c|}{$\sigma_0$ (nV$\cdot$s)}  & \multicolumn{2}{|c|}{SNR}\\
   (W) & (V) & copper cable & PoF & copper cable & PoF & copper cable & PoF \\
    \hline \hline
    0.84 & 45.00 $\pm$ 0.25 
    & 5.75 $\pm$ 0.04 
    & 6.04 $\pm$ 0.04  
    & 0.73 $\pm$ 0.04 
    & 0.80 $\pm$ 0.04
    & 7.83 $\pm$ 0.80 
    & 7.52 $\pm$ 0.19 \\  
    \hline
    0.93 & 46.00 $\pm$ 0.25
    & 7.41 $\pm$ 0.04   
    & 7.31 $\pm$ 0.04 
    & 0.69 $\pm$ 0.01 
    & 0.78 $\pm$ 0.02 
    & 10.66 $\pm$ 0.25 
    & 9.41 $\pm$ 0.26 \\
    \hline
    1.00 & 47.00 $\pm$ 0.26 
    & 8.95 $\pm$ 0.38 
    & 9.05 $\pm$ 0.05
    & 0.69 $\pm$ 0.01 
    & 0.82 $\pm$ 0.02 
    & 13.00$\pm$ 0.31
    & 11.07$\pm$ 0.32 \\
    \hline
    \end{tabular}
    }
\end{center}    
    \caption{Relevant parameters for the SNR calculation, measured for 20 Hamamatsu SiPMs, biased with standard copper cable and the Cryo-PoF system at three different bias voltages \cite{cryo_pof_creta}. The errors on the SiPM bias voltage refer to the voltage measured with the PoF system.}
    \label{tab::risultati}
\end{table}
\noindent
As shown in Table~\ref{tab::risultati}~\cite{cryo_pof_creta}, the performance of the PoF system is comparable to that of the copper-cable setup. 
Residual noise from the DC-DC converter is likely responsible for a slight degradation of the Cryo-PoF performance.

\begin{figure}[htbp]
\centering
\includegraphics[keepaspectratio=true,scale=0.12]{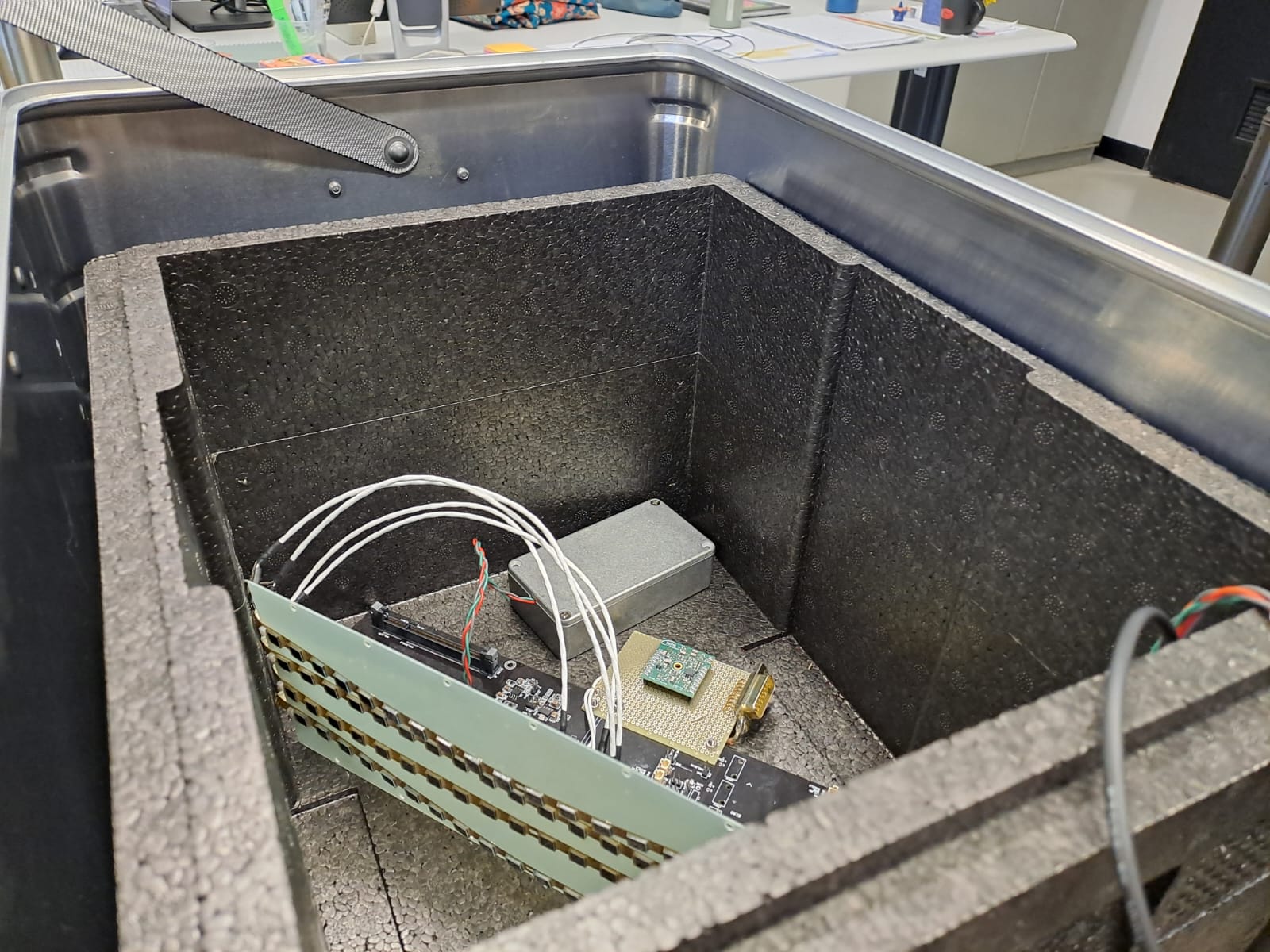}
\caption{Picture of the Cryo-PoF setup to power 80 FBK SiPMs.}
\label{fig::scatola_80}
\end{figure}

\begin{figure} [htpb]
\centering
\includegraphics[keepaspectratio=true,scale=0.26]{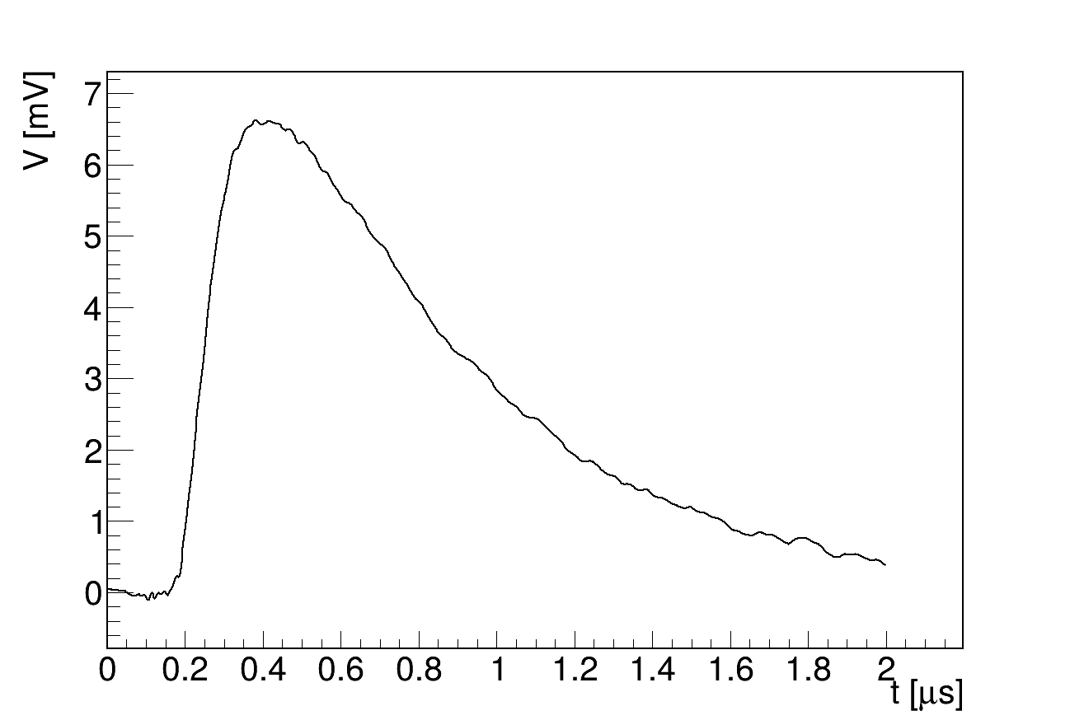}
\includegraphics[keepaspectratio=true,scale=0.26]
{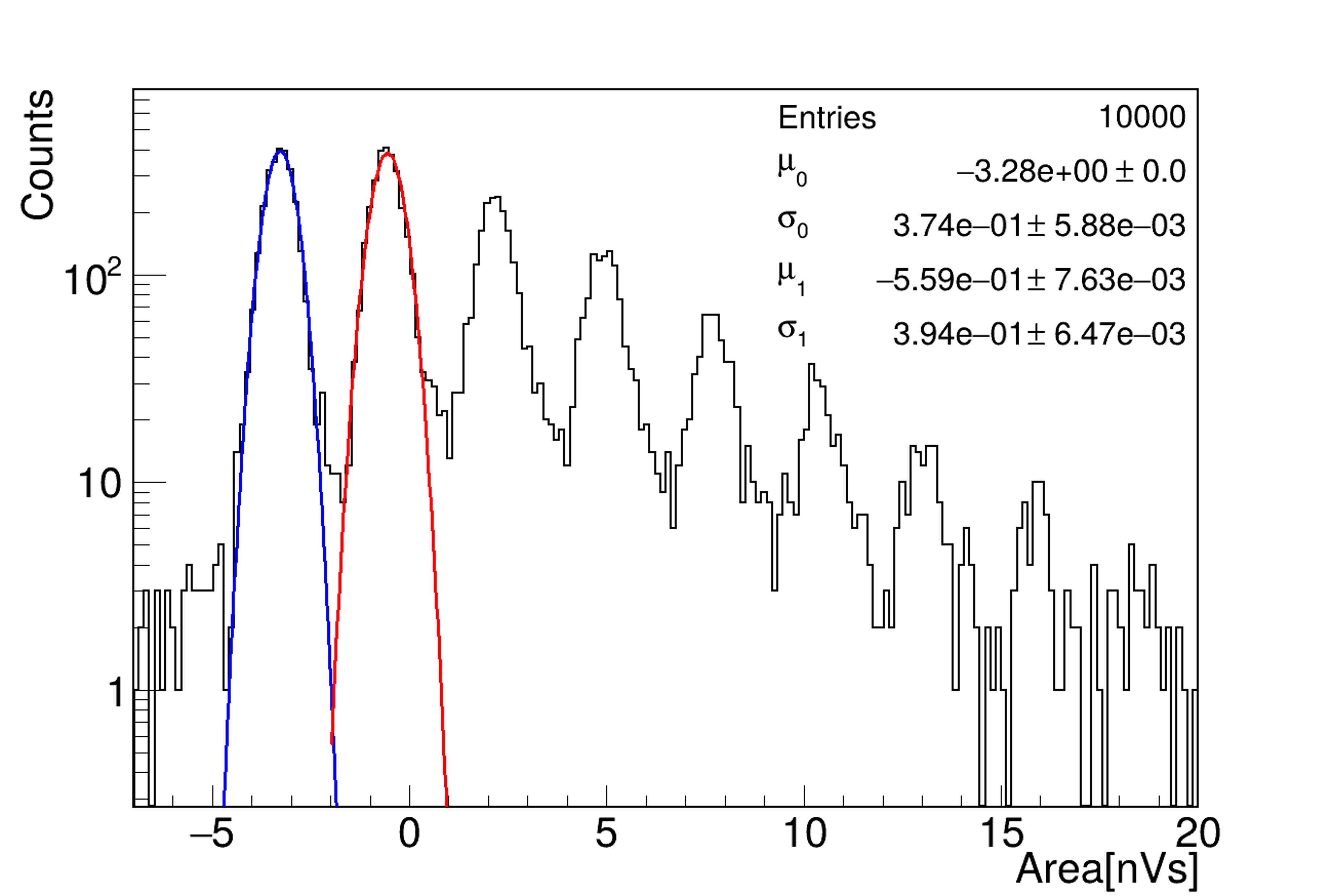}
\caption{Mean single-photoelectron (p.e.) waveform (left) and multi-p.e. spectra (right) obtained from 80 FBK SiPMs powered by the Cryo-PoF system. Measurements were taken at $V = 31.6$~V, corresponding to a laser power of $\sim$1.2~W. The SiPMs were illuminated with LED light, adjusted to yield approximately one or two p.e. detected per SiPM. The histogram shows the integrals (time window of 900~ns) of $10^4$ waveforms. The plot is fitted with a linear combination of two Gaussians, corresponding to 0 (blue) and 1 (red) recorded photoelectrons.}
\label{fig::ris_80}
\end{figure}

To mimic DUNE VD conditions, where one acquisition channel has 80 SiPMs, we tested the performance of the Cryo-PoF line with 80 FBK SiPMs (see Figure~\ref{fig::scatola_80}), using the same set-up and procedures described above. 
Tests were again performed at three different bias voltages: 30.6 V, 31.6 V, and 34.1 V, corresponding to a photon detection efficiency (PDE) of 40\%, 45\%, and 50\%, respectively \cite{FBK_article}. 
For each bias voltage, 10$^4$ waveforms were recorded. 
As an example, the mean single photoelectron waveform and the spectra obtained at 31.6 V are shown in Figure~\ref{fig::ris_80}.
The results (see Table~\ref{tab::risultati_80})~\cite{cryo_pof_elba} show a Signal to Noise Ratio exceeding 5 even at the lowest voltage, with the single photoelectron peak clearly distinguishable above the noise.
\begin{table}[htpb]
\begin{center}
    \begin{tabular}{|c|c|c|c|c|}
    \hline
    P laser (W) & SiPM bias (V) & $\mu_1 - \mu_0$ (nV$\cdot$s) & $\sigma_0$ (nV$\cdot$s) & SNR \\ 
    \hline\hline
    0.75 & 30.60 $\pm$ 0.25
    & 2.24 $\pm$ 0.01  
    & 0.37 $\pm$ 0.01   
    & 6.03 $\pm$ 0.12  \\  
    \hline
    0.93 & 31.60  $\pm$ 0.25
    & 2.72 $\pm$ 0.02  
    & 0.37 $\pm$ 0.06  
    & 7.27 $\pm$ 0.15 \\
    \hline
    1.38 & 34.10 $\pm$ 0.25
    & 4.34 $\pm$ 0.02 
    & 0.39 $\pm$ 0.06 
    & 11.27 $\pm$ 0.21 \\
    \hline
    \end{tabular}
\end{center}    
    \caption{Relevant parameters for the SNR calculation,  measured using the Cryo-PoF system with 80 FBK SiPMs, at three different bias voltages. }
    \label{tab::risultati_80}
\end{table}

\section{Conclusions}
\label{sec::conclusion}

The Power over Fiber (PoF) technology is intended to replace standard copper cables with optical fibers to power sensors in physical experiments, particularly when operating conditions are prohibitive, such as cryogenic environments or high-voltage surfaces. The technology uses laser light transmitted through an optical fiber to a photovoltaic power converter, which generates the required electrical power.\\
The Cryo-PoF project is developed within this framework, aiming to power both photosensors (SiPMs) and their cold electronics using a single PoF line, while tuning the SiPM voltage bias as a function of the laser power. 
Characterization of commercial devices (GaAs laser source and OPC) has shown very promising results for using this technology to power photosensors at cryogenic temperatures.\\
Our developed electronics can power both the SiPMs and the cold amplifier with a single PoF line, adjusting the SiPM bias voltage by tuning the laser power. Comparing the performance of SiPMs at different bias voltages, with and without PoF, in terms of Signal-to-Noise Ratio (SNR), we demonstrate that the Cryo-PoF system is competitive with standard copper cables. 
In addition, it offers bias tunability over a wide operating range by exploiting, through a custom DC-DC converter, the possibility to control the OPC output by means of the laser power.
During this measurement campaign, we also demonstrated the reliability of the tuning mechanism and its potential use in different voltage ranges.
%During this measurement campaign, we also validated the stability of this region over time and the reliability of the tuning mechanism.\\
The OPC efficiency of approximately 29\% at 4.6 K opens the possibility of new applications for this technology, such as in rare-event physics, quantum computing, and cryogenic setups that require ultra-low induced noise.
Cryo-PoF achieved very promising results; future developments for using PoF at cryogenic temperatures need more studies in order to investigate the stability, the degradation over time and the removal of induced noise.

\section{Acknowledgments}
%\noindent
The Cryo-PoF project was funded by the Istituto Nazionale di Fisica Nucleare (Italy) through the Grant "CSN5 Young Researcher Grant 2021" (Announcement n. 23246). 
This work was supported (in part) by projects of the Italian Ministry of Research (MUR): the PRIN2020 project "Photon detection in Extreme Environments for Fundamental and Applied Physics" (Grant No. PRIN 20208XN9TZ) and the PRIN2022 project "Beyond Liquid Argon (BeLAr)" (Grant No. PRIN 2022STFALX).
The authors would like to thank William Pellico (FNAL) and the FNAL and BNL DUNE groups for their support and valuable suggestions. 
Special thanks go to Mario Zannoni for providing instrumentation and support for testing the Cryo-PoF system at temperatures below 77 K. Finally, we are grateful to Broadcom for their guidance and to the mechanical workshop of INFN Milano Bicocca (G. Ceruti, R. Gaigher) for their help in realizing the measurement setup.

% Bibliography
%% [B] Manual formatting (see below)
%% (i) We suggest to always provide author, title and journal data or doi:
%% in short all the informations that clearly identify a document.
%% (ii) please avoid comments such as "For a review'', "For some examples",
%% "and references therein" or move them in the text. In general, please leave only references in the bibliography and move all
%% accessory text in footnotes.
%% (iii) Also, please have only one work for each \bibitem.

%\begin{thebibliography}{99}

%\bibitem{a}
%Author,
%\emph{Title},
%\emph{J. Abbrev.} {\bf vol} (year) pg.

%\bibitem{b}
%Author,
%\emph{Title},
%arxiv:1234.5678.

%\bibitem{c}
%Author,
%\emph{Title},
%Publisher (year).

%\end{thebibliography}
\end{document}